\documentclass[preprintnumbers, floatfix, onecolumn, preprintnumbers, letterpaper, superscriptaddress,nofootinbib]{revtex4}
\usepackage{amsfonts}
\usepackage{amsmath}
\usepackage{amssymb,epsf}
\usepackage{latexsym}
\usepackage{graphicx,epsfig}
\usepackage{amssymb}
\usepackage{subfigure}
\usepackage{pdflscape}
\usepackage{dcolumn}
\usepackage{bm}
\usepackage{color}
\usepackage[colorlinks=true,citecolor=blue,linkcolor=blue,urlcolor=black]{hyperref}

\begin{document}

\title{Generalized Mattig's relation in Brans-Dicke-Rastall gravity}
\author{Ines G. Salako}
\email{inessalako@gmail.com}
\affiliation{Institut de Math\'ematiques et de Sciences Physiques (IMSP), Universit\'e de
Porto-Novo, 01 BP 613 Porto-Novo, B\'enin}
\affiliation{D\'epartement de Physique - Universit\'e d'Agriculture de K\'etou, BP 13
K\'etou, B\'enin}
\author{M. J. S. Houndjo}
\email{sthoundjo@yahoo.fr}
\affiliation{Institut de Math\'ematiques et de Sciences Physiques (IMSP), Universit\'e de
Porto-Novo, 01 BP 613 Porto-Novo, B\'enin}
\affiliation{Facult\'e des Sciences et Techniques de Natitingou - Universit\'e de
Natitingou - B\'enin}
\author{Abdul Jawad}
\email{jawadab181@yahoo.com; abduljawad@ciitlahore.edu.pk}
\affiliation{Department of Mathematics, COMSATS Institute of Information Technology,\\
Lahore-54000, Pakistan}

\begin{abstract}
The Geodesic Deviation Equation is being studied in
Brans-Dicke-Rastall gravity. We briefly discuss the
Brans-Dicke-Rastall gravity and then construct GDE for FLRW metric.
In this way, the obtained geodesic deviation equation will
correspond to the Brans-Dicke-Rastall gravity.  Eventually, we
solve numerically the null vector GDE to obtain  from Mattig
relation, the deviation vector $\eta(z)$ and observer area distance
$r_0(z)$ and compare the results with $\Lambda$CDM model.
\end{abstract}

\maketitle

\section{Introduction\label{Intro}}

One of the main differences between general relativity (and its
alternatives) and Newtonian theory is that the effects of matter
distribution as well as gravitation are encoded in geometry of space-time.
In these contexts, the space-time is being curved due to the presence of
matter fields and can be described by a pair $(\mathcal{M},$ $\mathbf{g})$
where $\mathcal{M}$ and $\mathbf{g}$ correspond to a $d$-dimensional
manifold and the metric tensor on it respectively, while the curvature can
be represented by the Riemann tensor $\mathbf{R}$ \cite{Misner}. Therefore,
it is important to look for a relation which associates the curvature of
space-time to a physically measurable quantity. This aim is reachable via
Geodesic Deviation Equation (GDE) \cite{Synge,Pirani,Ellis2} in which the
Riemann tensor is connected to the relative acceleration between two close
test particles \cite{Wald,Poisson}. In fact, this relative acceleratrion is
produced by a kind of tidal force which is the result of the tendency of two
neighbour free falling particles to approch or recede from one another under
influence of a space-dependent gravitational field \cite%
{Synge,Pirani,Ellis2,Wald,Poisson}. Furthermore, in addition to elegant
description and interesting visualization that GDE presents about space-time
structure, various crucial solutions such as Raychaudhuri equation and
Mattig relation can be found through the solution of GDE for timelike, null
and spacelike geodesic congruences. One can also obtain the relative
acceleration of two neighbor geodesics by using GDE \cite{Wald,Poisson}.

Besides general relativity (GR) which shows relative successes in
various field strength regimes \cite{Corda}, there are various
alternative theories which extend GR from different points of view.
Some of these modifications to GR are constructed by adding a scalar
field degree of freedom to it which are called scalar-tensor
gravities \cite{nonlin}. Among scalar-tensor gravities, Brans-Dicke
(BD) theory \cite{BD1} is one of the most impressive and physically
viable modifications to GR. The motivations of this theory that are
Mach's principle and Dirac's large number hypothesis are encoded by
employing a scalar field $\phi $ which is proportional to the
inverse of
gravitational constant $G$ and coupled nonminimally to gravitation \cite%
{Brans1}. There is also a coupling constant $\omega $ in BD theory. This
theory reproduces GR in the limit $\omega \rightarrow \infty $ and $\phi =$ $%
constant$ \cite{KN,JDB,JMA}. Moreover, this theory is equivalence to
some other alternatives of GR in particular $f(R)$ gravities that
provide us a good tool to gain better insights about these theories
\cite{f(R)}. Other important property of BD theory is that it
produces simple expanding solutions \cite{CM} for scalar field $\phi
(t)$ and scale factor $a(t)$ which are more compatible with solar
system experiments \cite{PMG,SP,AGR}. There are also many works on
interesting physical aspects of the BD theory \cite{sergei}.

As we pointed out before, various geometrical theories which are
alternatives to GR have been put forward since its beginning for
explaining the gravitational phenomenon \cite{BD1,Cartan1922,
Cartan1923, Rosen1971, Moffat:1994hv, Bekenstein:2004ne,
Rastall1973, Rastall1976}. The conservation law (${T^{\mu \nu
}}_{;\mu }=0$) is one of them which does not hold true in a curved
space-time in some of these theories. Among these theories, the
steady-state model is the pioneer non-conservative theory of gravity
\cite{Bondi1948, Hoyle1948} which has been developed by following
some ideas already presented by Jordan \cite{Jordan:1949zz}.
Following this idea, Rastall has introduced alternative version of
non-conservative theory of gravity which suggests that gravitational
equations can be obtained via modification of conservation laws
\cite{Rastall1973, Rastall1976}. Also, the breaking of the weak and
null energy conditions for the average value of the EMT of a quantum
field in curved space-time was first shown in \cite{zel1} for weak
gravitational fields and in \cite{zel2} for strong gravitational
fields in cosmology. %It is curious to remark that the phenomenon of
%particle creation in cosmology\cite{Gibbons:1977mu, Parker:1971pt,
%Ford:1986sy} also leads to a violation of the weak and null energy
%conditions.
Hence, Rastall's idea leads to the classical formulation
of the quantum phenomenon because the violation of the
energy-momentum conservation is linked with curvature. The idea of
Rastall has also been employed to extend BD gravity and give rise to
Brans-Dicke-Rastall (BDR) gravity in which the conservation law is
no longer respected \cite{Carames,Carames'}.
% The GDE has already been studied in $f(R)$ \cite{eti35,eti36}, $f(T)$ \cite%
% {eti37}, $f(R,\mathcal{T})$ \cite{etienne} and $f(T,\mathcal{T})$ \cite%
% {salako} theories.

In the present work, our main goal is to study the GDE in
the metric context of Brans-Dicke-Rastall gravity.
 We also numerically solve the Mattig relation which helps
in measuring the cosmological
distances.

\section{Brief review on Brans-Dicke-Rastall theory}

\label{sec1}

\subsection{Rastall theory \ \ \ \ }

As we mentioned in introduction, Rastall has introduced alternative version
of non-conservative theory of gravity which proposes that equations of
gravitation can be obtained by modification of conservation laws as follow
\cite{Rastall1973, Rastall1976}%
\begin{equation}
{T^{\mu \nu }}_{;\mu }=\kappa R^{;\nu },
\end{equation}%
where $\kappa ,~T^{\mu \nu }$ and $R$ indicate the energy-momentum tensor, a
coupling constant and Ricci scalar curvature, respectively. In view of weak
field limit, the usual expressions can be recovered. We can also write the
above equation in terms of trace energy-momentum ($T$) as follows
\begin{equation}
{T^{\mu \nu }}_{;\mu }=\bar{\kappa}T^{;\nu }\;,
\end{equation}%
where $\bar{\kappa}$ is a new constant. In addition, Rastall's modification
to Einstein equations takes the form
\begin{eqnarray}
R_{\mu \nu }-\frac{1}{2}g_{\mu \nu }R &=&8\pi G\left( T_{\mu \nu }-\frac{%
\lambda _{Ras}-1}{2}g_{\mu \nu }T\right) \;,  \label{eq2R} \\
{T^{\mu \nu }}_{;\mu } &=&\frac{\lambda _{Ras}-1}{2}T^{;\nu }\;,
\end{eqnarray}%
where $c=1$ and $\lambda _{Ras}$ appears as Rastall's parameter and for $%
\lambda _{Ras}=1$, we can obtain GR. There also exists a possibility of
obtaining the above equations from Lagrangian formulation \cite{Smalley1984}%
. As $T=0,~R=0$ for a radiative fluid which implies that the cosmological
evolution during the radiative phase is the same as in the standard
cosmological scenario. At the same time, a single fluid inflationary model
as described by a cosmological constant is the same as it would be in GR
case. Hence, Rastall cosmologies seem like a curious departure from standard
cosmological model and from the beginning of the matter dominated phase \cite%
{Batista:2010nq, Fabris:2011rm, Fabris:2011wz, Capone:2009xm}.

\subsection{Brans-Dicke gravity}

The action of BD theory (where scalar field and matter field are
non-minimally coupled) is displayed below \cite{jamil} {\
\begin{equation}
S=\int {\ d^{4}x\sqrt{-g}\left( \phi {R}-\frac{\omega }{\phi }g^{\mu \nu
}\partial _{\mu }\phi \partial _{\nu }\phi -V(\phi )+L_{m}\right) },
\label{act1}
\end{equation}%
} where $\phi ,~V(\phi )$ and $\omega $ represents the BD scalar field,
potential and dimensionless BD parameter, respectively. The variation of
action (\ref{act1}) with respect to metric tensor yields the following
gravitational field equations
\begin{equation}
R_{\mu \nu }-\frac{1}{2}g_{\mu \nu }R=\frac{1}{\phi }T_{\mu \nu }+\frac{%
\omega }{\phi ^{2}}\Big(\phi _{\mu }\phi _{\nu }-\frac{1}{2}g_{\mu
\nu }\phi ^{\alpha }\phi _{\alpha }\Big)+\frac{1}{\phi }[\phi _{\mu
;\nu }-g_{\mu \nu }\Box \phi ]-g_{\mu \nu }\frac{V(\phi )}{2\phi },
\label{feq}
\end{equation}%
while $T_{\mu \nu }$ (stress-energy tensor) can be defined as follows
\begin{equation}
T_{\mu \nu }=(\rho +p)u_{\mu }u_{\nu }+pg_{\mu \nu },  \label{2aa}
\end{equation}%
where $u^{\mu }$ represents the four-vector velocity of the fluid which
satisfies $u^{\mu }u_{\mu }=-1$. Also $\rho $ and $p$ corresponds to the
energy density and pressure of various kinds of matters. One can describe the
various eras of the universe through equation of state parameter such as
radiation, dust and dark energy etc. The Klein-Gordon equation (or the wave
equation) for the scalar field has the form
\begin{equation}
\Box \phi =\frac{T}{3+2\omega}+\frac{1}{2\omega +3}(\phi V_{,\phi
}-2V),  \label{phi}
\end{equation}%
where $\Box =\nabla ^{\mu }\nabla _{\mu }$ and $\nabla _{\mu }$ represents
covariant derivative.

The trace of these ``Einsteinian equations'' (\ref{feq}) reads:
\begin{equation}
R=-\biggr\{\frac{1}{\phi }T-\frac{\omega }{\phi ^{2}}\phi _{;\rho }\phi
^{;\rho }-3\frac{\Box \phi }{\phi }-2\frac{V(\phi )}{\phi }\biggl\}.
\label{ricciscalar1}
\end{equation}%
With the aid of this expression equation (\ref{feq}) can be rewritten as
\begin{eqnarray}
R_{\mu \nu } &=&\frac{1}{\phi }T_{\mu \nu }+\frac{\omega }{\phi ^{2}}\Big(%
\phi _{\mu }\phi _{\nu }-\frac{1}{2}g_{\mu \nu }\phi ^{\alpha }\phi _{\alpha
}\Big)+\frac{1}{\phi }[\phi _{\mu ;\nu }-g_{\mu \nu }\Box \phi ]-g_{\mu \nu }%
\frac{V(\phi )}{2\phi }-\frac{1}{2}g_{\mu \nu }\biggr\{\frac{1}{\phi }T-%
\frac{\omega }{\phi ^{2}}\phi _{;\rho }\phi ^{;\rho }-\frac{3\Box \phi }{%
\phi }-\frac{2V(\phi )}{\phi }\biggl\}  \notag  \label{ricci0} \\
&=&\frac{1}{\phi }T_{\mu \nu }+\frac{3\omega}{%
2\phi ^{2}}\phi _{\mu }\phi _{\nu }-\frac{\omega }{2\phi ^{2}}g_{\mu
\nu }\phi ^{\alpha }\phi _{\alpha }+\frac{1}{\phi }\phi _{\mu ;\nu
}+\frac{1}{2\phi }g_{\mu \nu }\Bigg(\Box \phi -V(\phi )-T\Bigg).
\end{eqnarray}
In the following, we will consider the potential $ V(\phi) $ null as
in \cite{Carames,Carames'}.

\subsection{ Brans-Dicke-Rastall gravity}

Let us generalize Rastall's version of the field equations to the
Brans-Dicke case. Following the original formulation in the context of GR, a
minimal modification implies:
\begin{equation}
R_{\mu \nu }-\frac{\lambda }{2}g_{\mu \nu }R=\frac{8\pi }{\phi }T_{\mu \nu }+%
\frac{\omega }{\phi ^{2}}\biggr\{\phi _{;\mu }\phi _{;\nu }-\frac{1}{2}%
g_{\mu \nu }\phi _{;\rho }\phi ^{;\rho }\biggl\}+\frac{1}{\phi }(\phi _{;\mu
;\nu }-g_{\mu \nu }\Box \phi ).  \label{fe}
\end{equation}%
It is important to remark that even if the structure of the right
hand side is the same as in the BD gravity, the whole equation
(\ref{fe}) can be derived from a Lagrangian only when $\lambda =1$.

The trace of these \textquotedblleft Einsteinian
equations\textquotedblright\ reads:
\begin{equation}
R=\frac{1}{1-2\lambda }\biggr\{\frac{8\pi }{\phi }T-\frac{\omega }{\phi ^{2}}%
\phi _{;\rho }\phi ^{;\rho }-3\frac{\Box \phi }{\phi }\biggl\}.
\label{ricciscalar2}
\end{equation}%
With the aid of this expression equation (\ref{fe}) can be rewritten as
\begin{eqnarray}
R_{\mu \nu } &=&\frac{8\pi }{\phi }\biggr\{T_{\mu \nu }-\frac{1-\lambda }{%
2(1-2\lambda )}g_{\mu \nu }T\biggl\}+  \notag  \label{ricci} \\
&+&\frac{\omega }{\phi ^{2}}\biggr\{\phi _{;\mu }\phi _{;\nu }+\frac{\lambda
}{2(1-2\lambda )}g_{\mu \nu }\phi _{;\rho }\phi ^{;\rho }\biggl\}+  \notag \\
&+&\frac{1}{\phi }\biggr\{\phi _{;\mu ;\nu }+\frac{(1+\lambda )}{%
2(1-2\lambda )}g_{\mu \nu }\Box \phi \biggl\}  \notag \\
&+&\frac{g_{\mu \nu }}{2(1-2\lambda )}\biggr\{\frac{8\pi }{\phi }T-\frac{%
\omega }{\phi ^{2}}\phi _{;\rho }\phi ^{;\rho }-3\frac{\Box \phi }{\phi }%
\biggl\}.
\end{eqnarray}%
The Bianchi identities lead to
\begin{equation}
\Box \phi =\frac{8\pi \lambda }{3\lambda -2(1-2\lambda )\omega }T-\frac{%
\omega (1-\lambda )}{3\lambda -2(1-2\lambda )\omega }\frac{\phi ^{;\rho
}\phi _{;\rho }}{\phi }.
\end{equation}

The complete set of equations is:
\begin{eqnarray}
{T^{\mu \nu }}_{;\mu } &=&\frac{(1-\lambda )\phi }{16\pi }R^{,\nu }, \\
R_{\mu \nu } &=&\frac{8\pi }{\phi }\biggr\{T_{\mu \nu }-\frac{1-\lambda }{%
2(1-2\lambda )}g_{\mu \nu }T\biggl\}+  \notag \\
&+&\frac{\omega }{\phi ^{2}}\biggr\{\phi _{;\mu }\phi _{;\nu }+\frac{\lambda
}{2(1-2\lambda )}g_{\mu \nu }\phi _{;\rho }\phi ^{;\rho }\biggl\}+  \notag \\
&+&\frac{1}{\phi }\biggr\{\phi _{;\mu ;\nu }+\frac{(1+\lambda )}{%
2(1-2\lambda )}g_{\mu \nu }\Box \phi \biggl\}  \notag \\
&+&\frac{g_{\mu \nu }}{2(1-2\lambda )}\biggr\{\frac{8\pi }{\phi }T-\frac{%
\omega }{\phi ^{2}}\phi _{;\rho }\phi ^{;\rho }-3\frac{\Box \phi }{\phi }%
\biggl\}, \\
\Box \phi &=&\frac{8\pi \lambda }{3\lambda -2(1-2\lambda )\omega }T-\frac{%
\omega (1-\lambda )}{3\lambda -2(1-2\lambda )\omega }\frac{\phi ^{;\rho
}\phi _{;\rho }}{\phi }.
\end{eqnarray}%
\begin{figure}[t]
\epsfxsize=7cm \centerline{\epsffile{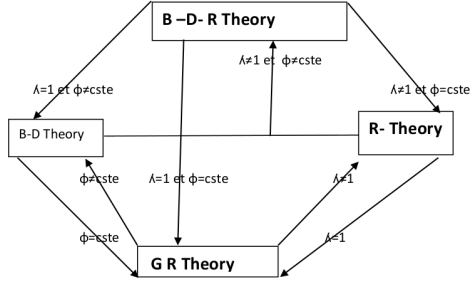}}
\caption{Brans-Dicke-Rastall theory} \label{fig1}
\end{figure}

Through Fig. (\ref{fig1}), we see how the BDR theory was
constructed. We can also note the passage from one theory to another
by changing the parameters $\phi$ and $\lambda$. Thus, when we take
$\lambda=1 $ in Rastall theory or when we take $\phi \sim 1/G $ in
BD theory or when we take $\phi \sim 1/G $and $\lambda=1$ in
Brans-Dicke-Rastall theory, the GR is recovered.

\section{Geodesic Deviation Equation}

\begin{figure}[t]
\epsfxsize=7cm \centerline{\epsffile{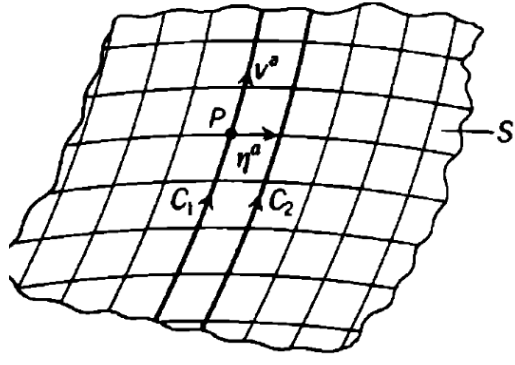}} \caption{Geodesic
deviation.} \label{fig2}
\end{figure}

Here, we will discuss the GDE by following \cite{Ellis2,Wald,Poisson}. Assuming that $%
\gamma _{0}$ and $\gamma _{1}$ are two neighboring geodesics along with
affine parameter $\nu $. There exists a family of interpolating geodesics $s$
which can be described as $x^{\alpha }(\nu ,s)$ (Fig. \ref{fig2}).
Consequently, the tangent to the geodesic is the vector field $V^{\alpha }=%
\frac{dx^{\alpha }}{d\nu }$, while $\eta ^{\alpha }=\frac{dx^{\alpha }}{ds}$
is the tangent vector field to the family $s$. However, the acceleration for
this vector field can be defined as follows \cite{Wald,Poisson}
\begin{equation}
\boxed{\frac{D^2 \eta^{\alpha}}{D \nu^2} = -
R_{\beta\gamma\delta}^{\alpha}V^{\beta}\eta^{\gamma}V^{\delta},}  \label{GDE}
\end{equation}%
which is also called GDE. In this equation, $\frac{D}{D\nu }$ indicates the
covariant derivative along the curve. The Riemann tensor can be written as
\cite{Wald,Ellis}
\begin{equation}
R_{\alpha \beta \gamma \delta }=C_{\alpha \beta \gamma \delta }+\frac{1}{2}%
\bigl(g_{\alpha \gamma }R_{\delta \beta }-g_{\alpha \delta }R_{\gamma \beta
}+g_{\beta \delta }R_{\gamma \alpha }-g_{\beta \gamma }R_{\delta \alpha }%
\bigr)-\frac{R}{6}\bigl(g_{\alpha \gamma }g_{\delta \beta }-g_{\alpha \delta
}g_{\gamma \beta }\bigr),  \label{RiemannC}
\end{equation}%
where $C_{\alpha \beta \gamma \delta }$ appears as the the Weyl tensor.

Next we defines the Friedman-Lama\^{\i}tre-Robertson-Walker (FLRW) universe
as follows
\begin{equation}
ds^{2}=-dt^{2}+a^{2}(t)\biggl[\frac{dr^{2}}{1-kr^{2}}+r^{2}d\theta
^{2}+r^{2}\sin ^{2}\theta d\varphi ^{2}\biggr],
\end{equation}%
where $a(t)$ represents the scale factor and $k$ corresponds to spatial
curvature of the universe. For this metric, Weyl tensor $C_{\alpha \beta
\gamma \delta }$ turns out to be zero. Through this paper we use the sign convention $(-, +, +, +)$
and geometrical units with $c = 1$. However, the trace of energy-momentum
tensor becomes
\begin{equation}
T=3p-\rho .
\end{equation}%
The Einstein field equations (with cosmological constant) in GR are
\begin{equation}
\boxed{R_{\alpha\beta} - \frac{1}{2}R g_{\alpha\beta} + \Lambda
g_{\alpha\beta} = \kappa T_{\alpha\beta}.}
\end{equation}%
Hence, Ricci scalar $R$ and Ricci tensor $R_{\alpha \beta }$ with the help
of above equation can be written as
\begin{gather}
R=\kappa (\rho -3p)+4\Lambda , \\
R_{\alpha \beta }=\kappa (\rho +p)u_{\alpha }u_{\beta }+\frac{1}{2}\bigl[%
\kappa (\rho -p)+2\Lambda \bigr]g_{\alpha \beta }.
\end{gather}%
With the help of these expressions, the right side of (\ref{GDE}) turns out
to be \cite{Ellis2}
\begin{equation}
R_{\beta \gamma \delta }^{\alpha }V^{\beta }\eta ^{\gamma }V^{\delta }=%
\biggl[\frac{1}{3}(\kappa \rho +\Lambda )\epsilon +\frac{1}{2}\kappa (\rho
+p)E^{2}\biggr]\eta ^{\alpha },  \label{Pirani1}
\end{equation}%
where $\epsilon =V^{\alpha }V_{\alpha }$ and $E=-V_{\alpha
}u^{\alpha }$. This equation is called \textit{Pirani equation}
\cite{Pirani}. The detailed study about GDE and some solutions for
spacelike, timelike and null congruences has given in \cite{Ellis2}.
Also, crucial results about cosmological distances has given in
\cite{Ellis}. This equation is well known \cite{Pirani} which
specify the spacelike components of the geodesic deviation vector
$\eta^\mu$ that describes the distance between two infinitesimally
close particles in free fall. In our study, we generalized these
results for BDR metric formalism.

\section{GDE in Brans-Dicke-Rastall gravity}

The expression (\ref{RiemannC}) in Brans-Dicke-Rastall gravity field
equations can be written as
\begin{eqnarray}
R_{\alpha \beta \gamma \delta } &=&C_{\alpha \beta \gamma \delta }  \notag \\
&+&\frac{8\pi }{\phi }\Biggl[(T_{\delta \beta }g_{\alpha \gamma }-T_{\gamma
\beta }g_{\alpha \delta }+T_{\gamma \alpha }g_{\beta \delta }-T_{\delta
\alpha }g_{\beta \gamma })\Biggr]+\biggl(\frac{8\pi \lambda \,T}{2\phi
(1-2\lambda )}+\frac{\omega (\lambda -1)\phi _{;\rho }\phi ^{;\rho }}{2\phi
^{2}(1-2\lambda )}\biggr)  \notag \\
&\times &\bigl(g_{\alpha \gamma }g_{\delta \beta }-g_{\alpha \delta
}g_{\gamma \beta }\bigr)+\frac{\omega }{\phi ^{2}}\bigl(g_{\alpha \gamma }%
\mathcal{D1}_{\delta \beta }-g_{\alpha \delta }\mathcal{D1}_{\gamma \beta
}+g_{\beta \delta }\mathcal{D1}_{\gamma \alpha }-g_{\beta \gamma }\mathcal{D1%
}_{\delta \alpha }\bigr)\phi  \notag \\
&+&\frac{1}{\phi }\bigl(g_{\alpha \gamma }\mathcal{D2}_{\delta \beta
}-g_{\alpha \delta }\mathcal{D2}_{\gamma \beta }+g_{\beta \delta }\mathcal{D2%
}_{\gamma \alpha }-g_{\beta \gamma }\mathcal{D2}_{\delta \alpha }\bigr)\phi
\notag \\
&+&\frac{\lambda -2}{2\phi (1-2\lambda )}\bigl(g_{\alpha \gamma }\mathcal{D3}%
_{\delta \beta }-g_{\alpha \delta }\mathcal{D3}_{\gamma \beta }+g_{\beta
\delta }\mathcal{D3}_{\gamma \alpha }-g_{\beta \gamma }\mathcal{D3}_{\delta
\alpha }\bigr)\phi  \notag \\
&+&\frac{\omega (\lambda -1)}{2\phi ^{2}(1-2\lambda )}\bigl(g_{\alpha \gamma
}\mathcal{D4}_{\delta \beta }-g_{\alpha \delta }\mathcal{D4}_{\gamma \beta
}+g_{\beta \delta }\mathcal{D4}_{\gamma \alpha }-g_{\beta \gamma }\mathcal{D4%
}_{\delta \alpha }\bigr)\phi  \notag \\
&-&\frac{1}{6(1-2\lambda )}\biggr\{\frac{8\pi }{\phi }T-\frac{\omega }{\phi
^{2}}\phi _{;\rho }\phi ^{;\rho }-3\frac{\Box \phi }{\phi }\biggl\}\bigl(%
g_{\alpha \gamma }g_{\delta \beta }-g_{\alpha \delta }g_{\gamma \beta }\bigr)%
.
\end{eqnarray}%
where we defined the operators
\begin{equation*}
\boxed{\mathcal{D1}_{\alpha\beta} \phi \equiv g_{\alpha\beta}
\phi_{;\alpha}\phi_{;\beta} },\boxed{\mathcal{D2}_{\alpha\beta} \phi \equiv
g_{\alpha\beta} \phi_{;\alpha;\beta} },\boxed{\mathcal{D3}_{\alpha\beta}
\equiv g_{\alpha\beta}\square \phi.},\boxed{\mathcal{D4}_{\alpha\beta} \phi
\equiv g_{\alpha\beta} \phi_{;\alpha}\phi^{;\beta}.}
\end{equation*}%
By rising the first index and contracting with $V^{\beta }\eta ^{\gamma
}V^{\delta }$ of Riemann tensor, the GDE turns out to be
\begin{eqnarray}
R_{\beta \gamma \delta }^{\alpha }V^{\beta }\eta ^{\gamma }V^{\delta }
&=&C_{\beta \gamma \delta }^{\alpha }V^{\beta }\eta ^{\gamma }V^{\delta }
\notag  \label{GeneralGDE} \\
&+&\frac{8\pi }{\phi }\Biggl[(T_{\delta \beta }\delta _{\gamma }^{\alpha
}-T_{\gamma \beta }\delta _{\delta }^{\alpha }+T_{\gamma }^{\,\,\alpha
}g_{\beta \delta }-T_{\delta }^{\,\,\alpha }g_{\beta \gamma })\Biggr]%
V^{\beta }\eta ^{\gamma }V^{\delta }  \notag \\
&+&\biggl(\frac{8\pi \lambda \,T}{2\phi (1-2\lambda )}+\frac{\omega (\lambda
-1)\phi _{;\rho }\phi ^{;\rho }}{2\phi ^{2}(1-2\lambda )}\biggr)\bigl(\delta
_{\gamma }^{\alpha }g_{\delta \beta }-\delta _{\delta }^{\alpha }g_{\gamma
\beta }\bigr)\,V^{\beta }\eta ^{\gamma }V^{\delta }  \notag \\
&+&\frac{\omega }{\phi ^{2}}\bigl(\delta _{\gamma }^{\alpha }\mathcal{D1}%
_{\delta \beta }-\delta _{\delta }^{\alpha }\mathcal{D1}_{\gamma \beta
}+g_{\beta \delta }\mathcal{D1}_{\gamma }^{\,\,\alpha }-g_{\beta \gamma }%
\mathcal{D1}_{\delta }^{\,\,\alpha }\bigr)\phi \,V^{\beta }\eta ^{\gamma
}V^{\delta }  \notag \\
&+&\frac{1}{\phi }\bigl(\delta _{\gamma }^{\alpha }\mathcal{D2}_{\delta
\beta }-\delta _{\delta }^{\alpha }\mathcal{D2}_{\gamma \beta }+g_{\beta
\delta }\mathcal{D2}_{\gamma }^{\,\,\alpha }-g_{\beta \gamma }\mathcal{D2}%
_{\delta }^{\,\,\alpha }\bigr)\phi \,V^{\beta }\eta ^{\gamma }V^{\delta }
\notag \\
&+&\frac{\lambda -2}{2\phi (1-2\lambda )}\bigl(\delta _{\gamma }^{\alpha }%
\mathcal{D3}_{\delta \beta }-\delta _{\delta }^{\alpha }\mathcal{D3}_{\gamma
\beta }+g_{\beta \delta }\mathcal{D3}_{\gamma }^{\,\,\alpha }-g_{\beta
\gamma }\mathcal{D3}_{\delta }^{\,\,\alpha }\bigr)\phi \,V^{\beta }\eta
^{\gamma }V^{\delta }  \notag \\
&+&\frac{(\lambda -1)\omega }{2\phi ^{2}(1-2\lambda )}\bigl(\delta _{\gamma
}^{\alpha }\mathcal{D4}_{\delta \beta }-\delta _{\delta }^{\alpha }\mathcal{%
D4}_{\gamma \beta }+g_{\beta \delta }\mathcal{D4}_{\gamma }^{\,\,\alpha
}-g_{\beta \gamma }\mathcal{D4}_{\delta }^{\,\,\alpha }\bigr)\phi \,V^{\beta
}\eta ^{\gamma }V^{\delta }  \notag \\
&-&\frac{1}{6(1-2\lambda )}\biggr\{\frac{8\pi }{\phi }T-\frac{\omega }{\phi
^{2}}\phi _{;\rho }\phi ^{;\rho }-3\frac{\Box \phi }{\phi }\biggl\}\bigl(%
\delta _{\gamma }^{\alpha }g_{\delta \beta }-\delta _{\delta }^{\alpha
}g_{\gamma \beta }\bigr)V^{\beta }\eta ^{\gamma }V^{\delta }.
\end{eqnarray}

\subsection{GDE for FLRW universe}

Here, we find the GDE in Brans-Dicke-Rastall gravity for FLRW metric and we
also compare our results with GR in the limiting case where $f(R)=R-2\Lambda
$. For FLRW metric, we have
\begin{equation}
R=\frac{1}{1-2\lambda }\biggr\{\frac{8\pi }{\phi }(3p-\rho )-\frac{\omega }{%
\phi ^{2}}\phi _{;\rho }\phi ^{;\rho }-3\frac{\Box \phi }{\phi }\biggl\}.
\label{ScalarRWF}
\end{equation}%
\begin{eqnarray}
R_{\alpha \beta } &=&\frac{8\pi }{\phi }\biggr\{(\rho +p)u_{\alpha }u_{\beta
}+pg_{\alpha \beta }-\frac{1-\lambda }{2(1-2\lambda )}g_{\alpha \beta
}(3p-\rho )\biggl\}+  \notag  \label{RicciRWF} \\
&+&\frac{\omega }{\phi ^{2}}\biggr\{\phi _{;\alpha }\phi _{;\beta }+\frac{%
\lambda }{2(1-2\lambda )}g_{\alpha \beta }\phi _{;\rho }\phi ^{;\rho }%
\biggl\}+  \notag \\
&+&\frac{1}{\phi }\biggr\{\phi _{;\alpha ;\beta }+\frac{(1+\lambda )}{%
2(1-2\lambda )}g_{\alpha \beta }\Box \phi \biggl\}  \notag \\
&+&\frac{g_{\alpha \beta }}{2(1-2\lambda )}\biggr\{\frac{8\pi }{\phi }%
(3p-\rho )-\frac{\omega }{\phi ^{2}}\phi _{;\rho }\phi ^{;\rho }-3\frac{\Box
\phi }{\phi }\biggl\}.
\end{eqnarray}%
With the help of above expressions, the Riemann tensor becomes
\begin{eqnarray}
R_{\alpha \beta \gamma \delta } &=&\frac{8\pi }{\phi }\Biggl[(\rho +p)\bigl(%
u_{\delta }u_{\beta }g_{\alpha \gamma }-u_{\gamma }u_{\beta }g_{\alpha
\delta }+u_{\gamma }u_{\alpha }g_{\beta \delta }-u_{\delta }u_{\alpha
}g_{\beta \gamma }\bigr)\Biggr]  \notag \\
&+&\Biggl[\biggl(\frac{8\pi \lambda \,(3p-\rho )}{2\phi (1-2\lambda )}+\frac{%
\omega (\lambda -1)\phi _{;\rho }\phi ^{;\rho }}{2\phi ^{2}(1-2\lambda )}%
\biggr)-\frac{1}{6(1-2\lambda )}\biggr\{\frac{8\pi }{\phi }(3p-\rho )-\frac{%
\omega }{\phi ^{2}}\phi _{;\rho }\phi ^{;\rho }-3\frac{\Box \phi }{\phi }%
\biggl\}\Biggr]\bigl(g_{\alpha \gamma }g_{\delta \beta }-g_{\alpha \delta
}g_{\gamma \beta }\bigr)  \notag \\
&+&\frac{\omega }{\phi ^{2}}\biggl(g_{\alpha \gamma }\mathcal{D1}_{\delta
\beta }-g_{\alpha \delta }\mathcal{D1}_{\gamma \beta }+g_{\beta \delta }%
\mathcal{D1}_{\gamma \alpha }-g_{\beta \gamma }\mathcal{D1}_{\delta \alpha }%
\biggr)\phi  \notag \\
&+&\frac{1}{\phi }\biggl(g_{\alpha \gamma }\mathcal{D2}_{\delta \beta
}-g_{\alpha \delta }\mathcal{D2}_{\gamma \beta }+g_{\beta \delta }\mathcal{D2%
}_{\gamma \alpha }-g_{\beta \gamma }\mathcal{D2}_{\delta \alpha }\biggr)\phi
\notag \\
&+&\frac{\lambda -2}{2\phi (1-2\lambda )}\biggl(g_{\alpha \gamma }\mathcal{D3%
}_{\delta \beta }-g_{\alpha \delta }\mathcal{D3}_{\gamma \beta }+g_{\beta
\delta }\mathcal{D3}_{\gamma \alpha }-g_{\beta \gamma }\mathcal{D3}_{\delta
\alpha }\biggr)\phi  \notag \\
&+&\frac{(\lambda -1)\omega }{2\phi ^{2}(1-2\lambda )}\biggl(g_{\alpha
\gamma }\mathcal{D4}_{\delta \beta }-g_{\alpha \delta }\mathcal{D4}_{\gamma
\beta }+g_{\beta \delta }\mathcal{D4}_{\gamma \alpha }-g_{\beta \gamma }%
\mathcal{D4}_{\delta \alpha }\biggr)\phi .
\end{eqnarray}%
For normalized vector field $V^{\alpha }$ we have $V^{\alpha }V_{\alpha
}=\epsilon $ and
\begin{eqnarray}
R_{\alpha \beta \gamma \delta }V^{\beta }V^{\delta } &=&\frac{8\pi }{\phi }%
\Biggl[(\rho +p)\bigl(g_{\alpha \gamma }(u_{\beta }V^{\beta
})^{2}-2(u_{\beta }V^{\beta })V_{(\alpha }u_{\gamma )}+\epsilon u_{\alpha
}u_{\gamma }\bigr)\Biggr]  \notag \\
&+&\Biggl[\biggl(\frac{8\pi \lambda \,(3p-\rho )}{2\phi (1-2\lambda )}+\frac{%
\omega (\lambda -1)\phi _{;\rho }\phi ^{;\rho }}{2\phi ^{2}(1-2\lambda )}%
\biggr)+\frac{1}{6(1-2\lambda )}\biggr\{\frac{8\pi }{\phi }(3p-\rho )-\frac{%
\omega }{\phi ^{2}}\phi _{;\rho }\phi ^{;\rho }-3\frac{\Box \phi }{\phi }%
\biggl\}\Biggr]\bigl(\epsilon g_{\alpha \gamma }-V_{\alpha }V_{\gamma }\bigr)
\notag \\
&+&\frac{\omega }{\phi ^{2}}\biggl(g_{\alpha \gamma }\mathcal{D1}_{\delta
\beta }-g_{\alpha \delta }\mathcal{D1}_{\gamma \beta }+g_{\beta \delta }%
\mathcal{D1}_{\gamma \alpha }-g_{\beta \gamma }\mathcal{D1}_{\delta \alpha }%
\biggr)\phi \,V^{\beta }V^{\delta }  \notag \\
&+&\frac{1}{\phi }\biggl(g_{\alpha \gamma }\mathcal{D2}_{\delta \beta
}-g_{\alpha \delta }\mathcal{D2}_{\gamma \beta }+g_{\beta \delta }\mathcal{D2%
}_{\gamma \alpha }-g_{\beta \gamma }\mathcal{D2}_{\delta \alpha }\biggr)\phi
\,V^{\beta }V^{\delta }  \notag \\
&+&\frac{\lambda -2}{2\phi (1-2\lambda )}\biggl(g_{\alpha \gamma }\mathcal{D3%
}_{\delta \beta }-g_{\alpha \delta }\mathcal{D3}_{\gamma \beta }+g_{\beta
\delta }\mathcal{D3}_{\gamma \alpha }-g_{\beta \gamma }\mathcal{D3}_{\delta
\alpha }\biggr)\phi \,V^{\beta }V^{\delta }  \notag \\
&+&\frac{(\lambda -1)\omega }{2\phi ^{2}(1-2\lambda )}\biggl(g_{\alpha
\gamma }\mathcal{D4}_{\delta \beta }-g_{\alpha \delta }\mathcal{D4}_{\gamma
\beta }+g_{\beta \delta }\mathcal{D4}_{\gamma \alpha }-g_{\beta \gamma }%
\mathcal{D4}_{\delta \alpha }\biggr)\phi \,V^{\beta }V^{\delta }.
\end{eqnarray}%
By rising the first index and contracting with $\eta ^{\gamma }$, the
Riemann tensor gives
\begin{eqnarray}
R_{\beta \gamma \delta }^{\alpha }V^{\beta }\eta ^{\gamma }V^{\delta } &=&%
\frac{8\pi }{\phi }\Biggl[(\rho +p)\bigl((u_{\beta }V^{\beta })^{2}\eta
^{\alpha }-(u_{\beta }V^{\beta })V^{\alpha }(u_{\gamma }\eta ^{\gamma
})-(u_{\beta }V^{\beta })u^{\alpha }(V_{\gamma }\eta ^{\gamma })+\epsilon
u^{\alpha }u_{\gamma }\eta ^{\gamma }\bigr)\Biggr]  \notag \\
&+&\Biggl[\biggl(\frac{8\pi \lambda \,(3p-\rho )}{2\phi (1-2\lambda )}+\frac{%
\omega (\lambda -1)\phi _{;\rho }\phi ^{;\rho }}{2\phi ^{2}(1-2\lambda )}%
\biggr)-\frac{1}{6(1-2\lambda )}\biggr\{\frac{8\pi }{\phi }(3p-\rho )-\frac{%
\omega }{\phi ^{2}}\phi _{;\rho }\phi ^{;\rho }-3\frac{\Box \phi }{\phi }%
\biggl\}\Biggr]  \notag \\
&\times &\bigl(\epsilon \eta ^{\alpha }-V^{\alpha }(V_{\gamma }\eta ^{\gamma
})\bigr)+\frac{\omega }{\phi ^{2}}\bigl[(\delta _{\gamma }^{\alpha }\mathcal{%
D1}_{\delta \beta }-\delta _{\delta }^{\alpha }\mathcal{D1}_{\gamma \beta
}+g_{\beta \delta }\mathcal{D1}_{\gamma }^{\,\,\alpha }-g_{\beta \gamma }%
\mathcal{D1}_{\delta }^{\,\,\alpha })\phi \bigr]V^{\beta }V^{\delta }\eta
^{\gamma }  \notag \\
&+&\frac{1}{\phi }\bigl[(\delta _{\gamma }^{\alpha }\mathcal{D2}_{\delta
\beta }-\delta _{\delta }^{\alpha }\mathcal{D2}_{\gamma \beta }+g_{\beta
\delta }\mathcal{D2}_{\gamma }^{\,\,\alpha }-g_{\beta \gamma }\mathcal{D2}%
_{\delta }^{\,\,\alpha })\phi \bigr]V^{\beta }V^{\delta }\eta ^{\gamma }
\notag \\
&+&\frac{\lambda -2}{2\phi (1-2\lambda )}\bigl[(\delta _{\gamma }^{\alpha }%
\mathcal{D3}_{\delta \beta }-\delta _{\delta }^{\alpha }\mathcal{D3}_{\gamma
\beta }+g_{\beta \delta }\mathcal{D3}_{\gamma }^{\,\,\alpha }-g_{\beta
\gamma }\mathcal{D3}_{\delta }^{\,\,\alpha })\phi \bigr]V^{\beta }V^{\delta
}\eta ^{\gamma }  \notag \\
&+&\frac{(\lambda -1)\omega }{2\phi ^{2}(1-2\lambda )}\bigl[(\delta _{\gamma
}^{\alpha }\mathcal{D4}_{\delta \beta }-\delta _{\delta }^{\alpha }\mathcal{%
D4}_{\gamma \beta }+g_{\beta \delta }\mathcal{D4}_{\gamma }^{\,\,\alpha
}-g_{\beta \gamma }\mathcal{D4}_{\delta }^{\,\,\alpha })\phi \bigr]V^{\beta
}V^{\delta }\eta ^{\gamma },
\end{eqnarray}%
where $E=-V_{\alpha }u^{\alpha },\eta _{\alpha }u^{\alpha }=\eta _{\alpha
}V^{\alpha }=0$ \cite{Ellis2}. Also, the above expression reduces to
\begin{eqnarray}
R_{\beta \gamma \delta }^{\alpha }V^{\beta }\eta ^{\gamma }V^{\delta } &=&%
\frac{8\pi }{\phi }\Biggl[(\rho +p)E^{2}\Biggr]  \notag \\
&+&\Biggl[\biggl(\frac{8\pi \lambda \,(3p-\rho )}{2\phi (1-2\lambda )}+\frac{%
\omega (\lambda -1)\phi _{;\rho }\phi ^{;\rho }}{2\phi ^{2}(1-2\lambda )}%
\biggr)-\frac{1}{6(1-2\lambda )}\biggr\{\frac{8\pi }{\phi }(3p-\rho )-\frac{%
\omega }{\phi ^{2}}\phi _{;\rho }\phi ^{;\rho }-3\frac{\Box \phi }{\phi }%
\biggl\}\Biggr]\epsilon  \notag \\
&+&\frac{\omega }{\phi ^{2}}\biggl[(\delta _{\gamma }^{\alpha }\mathcal{D1}%
_{\delta \beta }-\delta _{\delta }^{\alpha }\mathcal{D1}_{\gamma \beta
}+g_{\beta \delta }\mathcal{D1}_{\gamma }^{\,\,\alpha }-g_{\beta \gamma }%
\mathcal{D1}_{\delta }^{\,\,\alpha })\phi \biggr]V^{\beta }V^{\delta }\eta
^{\gamma }  \notag \\
&+&\frac{1}{\phi }\biggl[(\delta _{\gamma }^{\alpha }\mathcal{D2}_{\delta
\beta }-\delta _{\delta }^{\alpha }\mathcal{D2}_{\gamma \beta }+g_{\beta
\delta }\mathcal{D2}_{\gamma }^{\,\,\alpha }-g_{\beta \gamma }\mathcal{D2}%
_{\delta }^{\,\,\alpha })\phi \biggr]V^{\beta }V^{\delta }\eta ^{\gamma }
\notag \\
&+&\frac{\lambda -2}{2\phi (1-2\lambda )}\biggl[(\delta _{\gamma }^{\alpha }%
\mathcal{D3}_{\delta \beta }-\delta _{\delta }^{\alpha }\mathcal{D3}_{\gamma
\beta }+g_{\beta \delta }\mathcal{D3}_{\gamma }^{\,\,\alpha }-g_{\beta
\gamma }\mathcal{D3}_{\delta }^{\,\,\alpha })\phi \biggr]V^{\beta }V^{\delta
}\eta ^{\gamma }  \notag \\
&+&\frac{(\lambda -1)\omega }{2\phi ^{2}(1-2\lambda )}\biggl[(\delta
_{\gamma }^{\alpha }\mathcal{D4}_{\delta \beta }-\delta _{\delta }^{\alpha }%
\mathcal{D4}_{\gamma \beta }+g_{\beta \delta }\mathcal{D4}_{\gamma
}^{\,\,\alpha }-g_{\beta \gamma }\mathcal{D4}_{\delta }^{\,\,\alpha })\phi %
\biggr]V^{\beta }V^{\delta }\eta ^{\gamma }  \label{Riemann1}
\end{eqnarray}%
where $H\equiv \frac{\dot{a}}{a}$ is the Hubble parameter and $\phi $ is
only a function of time. Hence, the non-vanishing operators are
\begin{equation}
\square \phi =\ddot{\phi},\text{ \ \ \ \ }\mathcal{D1}_{00}=\dot{\phi}^{2},%
\text{ \ \ \ \ }\mathcal{D2}_{00}=\ddot{\phi},\text{ \ \ \ \ }\mathcal{D3}%
_{00}=\ddot{\phi},\text{ \ \ \ \ }\mathcal{D4}_{00}=\dot{\phi}^{2}
\label{square}
\end{equation}%
By make using of (\ref{square}), $R_{\beta \gamma \delta }^{\alpha }V^{\beta
}\eta ^{\gamma }V^{\delta }$ becomes
\begin{eqnarray}
R_{\beta \gamma \delta }^{\alpha }V^{\beta }\eta ^{\gamma }V^{\delta } &=&%
\Biggl[\frac{\omega \dot{\phi}^{2}+\phi (8\pi (p+\rho )-H\dot{\phi}+\ddot{%
\phi}}{2\phi ^{2}}E^{2}  \notag  \label{Pirani2} \\
&+&\epsilon \biggl(\frac{(-2+3\lambda )\omega \dot{\phi}^{2}+\phi (8\pi
(3p(-1+\lambda )+(-1+3\lambda )\rho )-3H(1+\lambda )\dot{\phi}+33(-1+\lambda
)\ddot{\phi})}{6(-1+2\lambda )\phi ^{2}}\biggr)\Biggr]\eta ^{\alpha }.
\end{eqnarray}%
This equation is the generalization of the Pirani equation for the
BDR metric formalism. Note that when $ \phi \sim 1/G$ and
$\lambda=1$, the previous equation reduces to (\ref{Pirani1}).
Finally, we can write the GDE (\ref{GDE}) in BDR gravity as the
following form
\begin{eqnarray}
\frac{D^{2}\eta ^{\alpha }}{D\nu ^{2}} &=&-\Biggl[\frac{\omega \dot{\phi}%
^{2}+\phi (8\pi (p+\rho )-H\dot{\phi}+\ddot{\phi}}{2\phi ^{2}}E^{2}  \notag
\label{GDeFR} \\
&+&\epsilon \biggl(\frac{(-2+3\lambda )\omega \dot{\phi}^{2}+\phi (8\pi
(3p(-1+\lambda )+(-1+3\lambda )\rho )-3H(1+\lambda )\dot{\phi}+33(-1+\lambda
)\ddot{\phi})}{6(-1+2\lambda )\phi ^{2}}\biggr)\Biggr]\eta ^{\alpha }.
\end{eqnarray}%
Thus the GDE produces variation only in the magnitude of the deviation
vector $\eta ^{\alpha }$, which also occurs in GR. This result is expected
FLRW universe. The GDE produces variation \textit{in the direction} of the
deviation vector in case of anisotropic universes (such as Bianchi I) \cite%
{Caceres}.

\subsection{GDE for fundamental observers}

As, $V^{\alpha }$ appears as a four-velocity of the fluid $u^{\alpha }$. For
temporal geodesics, we have $\epsilon =-1$ and for normalized vector field,
we have $E=1$. For these conditions, Eq.(\ref{Pirani2}) gives
\begin{equation}
\boxed{R_{\beta\gamma\delta}^{\alpha}u^{\beta}\eta^{\gamma}u^{\delta} =
\biggl[\frac{(-1 + 3 \lambda) \omega \dot{\phi}^2 + \phi (8 \pi (-2 \rho + 3
\lambda (p + \rho)) - 3 H (-2 + \lambda) \dot{\phi} + 3 \lambda
\ddot{\phi})}{6 (-1 + 2 \lambda) \phi^2} \biggr]\eta^{\alpha}.}
\label{GDEfR1}
\end{equation}%
If the deviation vector is $\eta _{\alpha }=\ell e_{\alpha }$, isotropy
implies
\begin{equation}
\frac{De^{\alpha }}{Dt}=0,
\end{equation}%
and
\begin{equation}
\frac{D^{2}\eta ^{\alpha }}{Dt^{2}}=\frac{d^{2}\ell }{dt^{2}}e^{\alpha }.
\end{equation}%
Using this result in GDE (\ref{GDE}) and (\ref{GDEfR1}) gives
\begin{equation}
\frac{d^{2}\ell }{dt^{2}}=-\biggl[\frac{(-1+3\lambda )\omega \dot{\phi}%
^{2}+\phi (8\pi (-2\rho +3\lambda (p+\rho ))-3H(-2+\lambda )\dot{\phi}%
+3\lambda \ddot{\phi})}{6(-1+2\lambda )\phi ^{2}}\biggr]\,\ell .
\end{equation}%
For $\ell =a(t)$, we have
\begin{equation}
\boxed{\frac{\ddot{a}}{a} = -\biggl[\frac{(-1 + 3 \lambda) \omega
\dot{\phi}^2 + \phi (8 \pi (-2 \rho + 3 \lambda (p + \rho)) - 3 H (-2 +
\lambda) \dot{\phi} + 3 \lambda \ddot{\phi})}{6 (-1 + 2 \lambda) \phi^2}
\biggr].}  \label{Raycha}
\end{equation}%
This is a particular case of the generalized Raychaudhuri equation given in
\cite{Rippl}. In standard form of the modified Friedmann equations, the
Raychaudhuri equation gives \cite{f(R)}%
\begin{eqnarray}
3H^{2} &=&\frac{8\pi }{\phi }\biggr\{\frac{(1-3\lambda )\rho}{%
2(1-2\lambda )}+\frac{3p(1-\lambda )}{2(1-2\lambda )} \biggl\}%
 + \omega \biggr[\frac{2-3\lambda }{2(1-2\lambda )}\biggl]\biggr(\frac{\dot{%
\phi}}{\phi }\biggl)^{2}+\biggr[\frac{3(1-\lambda )}{2(1-2\lambda )}\frac{%
\ddot{\phi}}{\phi }  \notag \\
&+&\frac{3(1+\lambda )}{2(1-2\lambda )}H\frac{\dot{\phi}}{\phi }\biggl],
\label{ModFried1} \\
2\dot{H}+3H^{2} &=&-\frac{8\pi }{\phi }\biggr\{\frac{\rho(1-\lambda)
-(1+\lambda )p}{2(1-2\lambda )}\biggl\} +\omega \frac{%
\lambda }{2(1-2\lambda )}\biggr(\frac{\dot{\phi}}{\phi }\biggl)^{2}+\frac{%
1+\lambda }{2(1-2\lambda )}\frac{\ddot{\phi}}{\phi }+\frac{5-\lambda }{%
2(1-2\lambda )}H\frac{\dot{\phi}}{\phi }.  \notag \\
&&
\end{eqnarray}

\subsection{GDE for null vector fields}

Here, we assume the GDE for null vector fields past directed for which $%
V^{\alpha }=k^{\alpha }$ and $k_{\alpha }k^{\alpha }=0$. Under these
conditions, Eq.(\ref{Pirani2}) becomes
\begin{equation}
\boxed{R_{\beta\gamma\delta}^{\alpha}k^{\beta}\eta^{\gamma}k^{\delta} =
\frac{\omega \dot{\phi}^2 + \phi (8 \pi (p + \rho) - H \dot{\phi} +
\ddot{\phi}}{2\phi^2} E^2\,\eta^{\alpha},}  \label{RicciFoc}
\end{equation}%
which can be described as the \textit{Ricci focusing} in $f(R)$ gravity. By
choosing $\eta ^{\alpha }=\eta e^{\alpha }$, $e_{\alpha }e^{\alpha }=1$, $%
e_{\alpha }u^{\alpha }=e_{\alpha }k^{\alpha }=0$ and aligned base parallel
propagated $\frac{De^{\alpha }}{D\nu }=k^{\beta }\nabla _{\beta }e^{\alpha
}=0$, the GDE (\ref{GDeFR}) turns out to be
\begin{equation}
\frac{d^{2}\eta }{d\nu ^{2}}=-\frac{\omega \dot{\phi}^{2}+\phi (8\pi (p+\rho
)-H\dot{\phi}+\ddot{\phi}}{2\phi ^{2}}E^{2}\,\eta .  \label{GDE3}
\end{equation}%
In the case of GR discussed in \cite{Ellis2}, all families of
past-directed null geodesics experience focusing, provided $\kappa
(\rho +p)>0$, and for a fluid with equation of state $p=-\rho $
(cosmological constant) there is no influence in the focusing
\cite{Ellis2}. From (\ref{GDE3}) the focusing condition for BDR
gravity is
\begin{equation}
\frac{\omega \dot{\phi}^{2}+\phi (8\pi (p+\rho )-H\dot{\phi}+\ddot{\phi}}{%
2\phi ^{2}}>0.
\end{equation}%
Eq. (\ref{GDE3}) can be written in terms of redshift parameter $z$ with the
help of following differential operators as
\begin{equation}
\frac{d}{d\nu }=\frac{dz}{d\nu }\frac{d}{dz},
\end{equation}%
\begin{equation}
\frac{d^{2}}{d\nu ^{2}}=\frac{dz}{d\nu }\frac{d}{dz}\biggl(\frac{d}{d\nu }%
\biggr)=\biggl(\frac{d\nu }{dz}\biggr)^{-2}\biggl[-\biggl(\frac{d\nu }{dz}%
\biggr)^{-1}\frac{d^{2}\nu }{dz^{2}}\frac{d}{dz}+\frac{d^{2}}{dz^{2}}\biggr].
\end{equation}%
For null geodesics, we have
\begin{equation}
(1+z)=\frac{a_{0}}{a}=\frac{E}{E_{0}}\quad \longrightarrow \quad \frac{dz}{%
1+z}=-\frac{da}{a},
\end{equation}%
where $a_{0}=1$ the present value of the scale factor. For past directed
case, we have
\begin{equation}
dz=(1+z)\frac{1}{a}\frac{da}{d\nu }\,d\nu =(1+z)\frac{\dot{a}}{a}E\,d\nu
=E_{0}H(1+z)^{2}\,d\nu ,\Rightarrow \frac{d\nu }{dz}=\frac{1}{E_{0}H(1+z)^{2}%
}
\end{equation}%
and
\begin{equation}
\frac{d^{2}\nu }{dz^{2}}=-\frac{1}{E_{0}H(1+z)^{3}}\biggl[\frac{1}{H}(1+z)%
\frac{dH}{dz}+2\biggr].
\end{equation}%
Also, we can write
\begin{equation}
\frac{dH}{dz}=\frac{d\nu }{dz}\frac{dt}{d\nu }\frac{dH}{dt}=-\frac{1}{H(1+z)}%
\frac{dH}{dt}.
\end{equation}%
Here, minus sign is due to past directed geodesic, when $z$ increases, $\nu $
decreases. Also, $\frac{dt}{d\nu }=E_{0}(1+z)$. The derivative of Hubble
parameter $H$ also yields
\begin{equation}
\dot{H}\equiv \frac{dH}{dt}=\frac{d}{dt}\frac{\dot{a}}{a}=\frac{\ddot{a}}{a}%
-H^{2},
\end{equation}%
The Raychaudhuri equation (\ref{Raycha}) yields
\begin{equation}
\dot{H}=\biggl[\frac{(-1+3\lambda )\omega \dot{\phi}^{2}+\phi (8\pi (-2\rho
+3\lambda (p+\rho ))-3H(-2+\lambda )\dot{\phi}+3\lambda \ddot{\phi})}{%
6(1-2\lambda )\phi ^{2}}\biggr]-H^{2},
\end{equation}%
and
\begin{equation}
\frac{d^{2}\nu }{dz^{2}}=-\frac{3}{E_{0}H(1+z)^{3}}\biggl[1+\frac{1}{3H^{2}}%
\biggl(\frac{(-1+3\lambda )\omega \dot{\phi}^{2}+\phi (8\pi (-2\rho
+3\lambda (p+\rho ))-3H(-2+\lambda )\dot{\phi}+3\lambda \ddot{\phi})}{%
6(-1+2\lambda )\phi ^{2}}\biggr)\biggr].
\end{equation}%
Finally, the operator $\frac{d^{2}\eta }{d\nu ^{2}}$ is
\begin{equation}
\frac{d^{2}\eta }{d\nu ^{2}}=\bigl(EH(1+z)\bigr)^{2}\Biggl\{\frac{d^{2}\eta
}{dz^{2}}+\frac{3}{(1+z)}\biggl[1+\frac{1}{3H^{2}}\biggl(\frac{(-1+3\lambda
)\omega \dot{\phi}^{2}+\phi (8\pi (-2\rho +3\lambda (p+\rho ))-3H(-2+\lambda
)\dot{\phi}+3\lambda \ddot{\phi})}{6(-1+2\lambda )\phi ^{2}}\biggr)\biggr]%
\frac{d\eta }{dz}\Biggr\},
\end{equation}%
and the GDE (\ref{GDE3}) reduces to
\begin{eqnarray}
\frac{d^{2}\eta }{dz^{2}} &+&\frac{3}{(1+z)}\Biggl[\frac{18H^{2}(-1+2\lambda
)\phi ^{2}+(-1+3\lambda )\omega \dot{\phi}^{2}+\phi (8\pi (-2\rho +3\lambda
(p+\rho ))-3H(-2+\lambda )\dot{\phi}+3\lambda \ddot{\phi})}{%
18H^{2}(-1+2\lambda )\phi ^{2}}\Biggr]\,\frac{d\eta }{dz}  \notag
\label{salako57} \\
&+&\frac{\omega \dot{\phi}^{2}+\phi (8\pi (p+\rho )-H\dot{\phi}+\ddot{\phi}}{%
2\phi ^{2}\,H^{2}(1+z)^{2}}\eta =0.
\end{eqnarray}%
This equation is so complicated and is difficult to solve
analytically. In order to solve it, it is possible to rewrite as a
differential equation containing only the unknown $H(z)$ and $f(z)$
and their derivatives with respect to $z$. By taking the
contributions of radiation dominated era, then the energy density
$\rho $ and the pressure $p$ turns out to be
\begin{equation}
\frac{8\pi }{\phi }\rho =3H_{0}^{2}\Omega _{m0}(1+z)^{3}+3H_{0}^{2}\Omega
_{r0}(1+z)^{4},\qquad \frac{8\pi }{\phi }p=H_{0}^{2}\Omega _{r0}(1+z)^{4},
\end{equation}%
where $p=p_{r}=\frac{1}{3}\rho _{r}$. The GDE takes the following form
\begin{equation}
\frac{d^{2}\eta }{dz^{2}}+\mathcal{P}(H,R,z)\frac{d\eta }{dz}+\mathcal{Q}%
(H,R,z)\eta =0,  \label{MattigGen}
\end{equation}%
with
\begin{eqnarray}
\mathcal{P}(H,R,z) &=&\frac{3}{(1+z)}\left\{ \frac{18H_{0}^{2}\biggr \{%
\biggr(\Omega _{m0}(1+z)^{3}+\Omega _{r0}(1+z)^{4}\biggl)\biggr[\frac{%
1-3\lambda }{2(1-2\lambda )}\biggl]+\Omega _{DE}\biggl\}(-1+2\lambda )\phi
^{2}+(-1+3\lambda )\omega \dot{\phi}^{2}}{18H_{0}^{2}\biggr \{\biggr(\Omega
_{m0}(1+z)^{3}+\Omega _{r0}(1+z)^{4}\biggl)\biggr[\frac{1-3\lambda }{%
2(1-2\lambda )}\biggl]+\Omega _{DE}\biggl\}(-1+2\lambda )\phi ^{2}}\right.
\notag \\
&&+\frac{-2\phi ^{2}3H_{0}^{2}\Omega _{m0}(1+z)^{3}+3H_{0}^{2}\Omega
_{r0}(1+z)^{4}+3\lambda \,\phi ^{2}3H_{0}^{2}\Omega
_{m0}(1+z)^{3}+4H_{0}^{2}\Omega _{r0}(1+z)^{4}}{18H_{0}^{2}\biggr \{\biggr(%
\Omega _{m0}(1+z)^{3}+\Omega _{r0}(1+z)^{4}\biggl)\biggr[\frac{1-3\lambda }{%
2(1-2\lambda )}\biggl]+\Omega _{DE}\biggl\}(-1+2\lambda )\phi ^{2}}  \notag
\\
&&\left. -\frac{3H(-2+\lambda )\dot{\phi}+3\lambda \ddot{\phi})}{18H_{0}^{2}%
\biggr \{\biggr(\Omega _{m0}(1+z)^{3}+\Omega _{r0}(1+z)^{4}\biggl)\biggr[%
\frac{1-3\lambda }{2(1-2\lambda )}\biggl]+\Omega _{DE}\biggl\}(-1+2\lambda
)\phi ^{2}}\right\} ,  \label{salako58}
\end{eqnarray}%
\begin{equation}
\mathcal{Q}(H,R,z)=\frac{\omega \dot{\phi}^{2}+\Big(3H_{0}^{2}\Omega
_{m0}(1+z)^{3}+4H_{0}^{2}\Omega _{r0}(1+z)^{4}\Big)-H\phi \,\dot{\phi}+\phi
\,\ddot{\phi}}{2\phi ^{2}\,H_{0}^{2}\biggr \{\biggr(\Omega
_{m0}(1+z)^{3}+\Omega _{r0}(1+z)^{4}\biggl)\biggr[\frac{1-3\lambda }{%
2(1-2\lambda )}\biggl]+\Omega _{DE}\biggl\}(1+z)^{2}},  \label{salako59}
\end{equation}%
and $H$ given by the modified field equations (\ref{ModFried1})
\begin{equation}
H^{2}=H_{0}^{2}\biggr \{\biggr(\Omega _{m0}(1+z)^{3}+\Omega _{r0}(1+z)^{4}%
\biggl)\biggr[\frac{1-3\lambda }{2(1-2\lambda )}\biggl]+\Omega _{DE}\biggl\},
\end{equation}%
where%
\begin{equation}
\boxed{\Omega_{DE} \equiv \frac{1}{3 H_0^2} \Bigg \{\Lambda +
\omega\biggr[\frac{2 - 3\lambda}{2(1 -
2\lambda)}\biggl]\biggr(\frac{\dot\phi}{\phi} \biggl)^2 + \biggr[\frac{3(1 -
\lambda)}{2(1 - 2\lambda)}\frac{\ddot\phi}{\phi} + \frac{3(1 + \lambda)}{2(1
- 2\lambda)}H\frac{\dot \phi}{\phi} +\frac{3(1 - \lambda)}{2(1 - 2\lambda)}p
\Bigg \}.}  \label{OmegaDE}
\end{equation}

At present, we will verify the consistency of our results with those
found in GR by taking the special case $\lambda \sim 1$, $\omega
\sim \infty $ and $\phi \sim \frac{1}{G}$. Therefore $\Omega _{DE}$
in Eq.(\ref{OmegaDE}) reduces to
\begin{equation}
\Omega _{DE}=\frac{\Lambda }{3H_{0}^{2}}\equiv \Omega _{\Lambda },
\label{salako62}
\end{equation}%
which allows us to rewrite the first Friedmann equation in GR as following
\begin{equation}
H^{2}=H_{0}^{2}[\Omega _{m0}(1+z)^{3}+\Omega _{r0}(1+z)^{4}+\Omega _{\Lambda
}].  \label{salako63}
\end{equation}

Thus, the expressions $P$ and $Q$ become
\begin{equation}
P(z)=\frac{\frac{7}{2}\Omega _{m0}(1+z)^{3}+4\Omega _{r0}(1+z)^{4}+2\Omega
_{\Lambda }}{(1+z)[\Omega _{m0}(1+z)^{3}+\Omega _{r0}(1+z)^{4}+\Omega
_{\Lambda }]},  \label{salako64}
\end{equation}%
\begin{equation}
Q(z)=\frac{3\Omega _{m0}(1+z)+4\Omega _{r0}(1+z)^{2}}{2[\Omega
_{m0}(1+z)^{3}+\Omega _{r0}(1+z)^{4}+\Omega _{\Lambda }]}.  \label{salako65}
\end{equation}%
Ultimately, the GDE for null vector fields becomes
\begin{widetext}
\begin{eqnarray}\label{salako66}
\frac{d^{2}\eta}{dz^{2}}+\frac{\frac{7}{2}\Omega_{m0}(1+z)^{3}+4\Omega_{r0}(1+z)^{4}+
2\Omega_{\Lambda}}{(1+z)[\Omega_{m0}(1+z)^{3}+\Omega_{r0}(1+z)^{4}+\Omega_{\Lambda}]}\frac{d\eta}{dz}+
\frac{3\Omega_{m0}(1+z)+4\Omega_{r0}(1+z)^{2}}{2(\Omega_{m0}(1+z)^{3}+\Omega_{r0}(1+z)^{4}+\Omega_{\Lambda})}\eta=0.
\end{eqnarray}
\end{widetext}
we note that to obtain  the GDE for null vector fields in GR
\cite{sch}, we have used $\Omega _{r0}+\Omega _{m0}=1$ which leads
to
\begin{widetext}
\begin{eqnarray}\label{salako67}
\frac{d^{2}\eta}{dz^{2}}+\frac{\frac{7}{2}\Omega_{m0}(1+z)^{3}+4\Omega_{r0}(1+z)^{4}}{(1+z)[\Omega_{m0}(1+z)^{3}+
\Omega_{r0}(1+z)^{4}]}\frac{d\eta}{dz}+\frac{3\Omega_{m0}(1+z)
+4\Omega_{r0}(1+z)^{2}}{2(\Omega_{m0}(1+z)^{3}+\Omega_{r0}(1+z)^{4})}\eta=0.
\end{eqnarray}
\end{widetext}

\subsection{Mattig's Equation}
Mattig's equation is one of the most important formulae in
observational cosmology and extragalactic astronomy which gives
relation between radial coordinate and redshift of a given source.
It depends on the cosmological model which is being used and is
needed to calculate luminosity distance in terms of redshift. In
BDR, the Mattig's equation can be interpreted by using
Eq.(\ref{MattigGen}) as follows:
\begin{equation}
r_{0}(z)|_{BDR}=\sqrt{\left\vert \frac{dA_{0}(z)}{d\Omega }\right\vert }=\left\vert
\frac{\eta_{BDR} (z^{\prime })|_{z}}{d\eta_{BDR} (z^{\prime })/d\ell |_{z^{\prime }=0}}%
\right\vert ,  \label{salako68}
\end{equation}%
where $A_{0}$ is the area of the object and also $\Omega $ is the
solid angle.  This equation (\ref{salako68}) could be
interpreted as a generalization of the Mattig relation in BDR
gravity because $\eta(z)$ is numerical solution which provide from
(\ref{MattigGen}). The equation (\ref{salako68}) can also be used
to generate the observer area distance $r_{0}(z)$ and the deviation
vector $\eta(z)$. Analytical expression for the observable area
distance for GR with no cosmological constant can be found in
\cite{sch}, whereas for more general scenarios numerical integration
is usually required. Equipped with the $d/d\ell
=E_{0}^{-1}(1+z)^{-1}d/d\nu =H(1+z)d/dz$ and setting the deviation
at $z=0$ to zero, clearly we have
\begin{equation}
r_{0}(z)|_{BDR}=\left\vert \frac{\eta_{BDR} (z)}{H(0)d\eta_{BDR} (z^{\prime })/dz^{\prime
}|_{z^{\prime }=0}}\right\vert ,  \label{salako69}
\end{equation}%
where $H(0)$ is the evaluated modified Friedmann equation
(\ref{salako62}) at $z=0$. In the next section, the numerical
analysis will also be performed in order to find the deviation
vector $\eta(z)$ and observer area distance $r_0(z)$ and compare the
results of BDR Jordan frame with those of $\Lambda$CDM model.

\subsection{Numerically solution of GDE for null vector fields in
BDR gravity}

In order to solve numerically the null vector GDE in BDR gravity, we
consider the model $\phi=\phi_0\,t^\sigma $ \cite{model1} where
$\sigma$ is constant parameter and $\phi_0$ is positive quantity
representing the present day values of the corresponding quantities.
In agreement with this model, the equations (\ref{salako58}),
(\ref{salako59}, (\ref{OmegaDE}) and (\ref{MattigGen}) can be
rewritten as follow
\begin{equation}
\frac{d^{2}\eta }{dz^{2}}+P\,\frac{d\eta }{dz}+Q\,\eta =0,
\label{MattigGen'}
\end{equation}%
\begin{equation}
\Omega _{DE}=\frac{1}{3H_{0}^{2}}\Bigg \{\Lambda +\omega \biggr[\frac{%
2-3\lambda }{2(1-2\lambda )}\biggl]\biggr(\frac{\dot{\phi}}{\phi }\biggl)%
^{2}+\biggr[\frac{3(1-\lambda )}{2(1-2\lambda )}\frac{\ddot{\phi}}{\phi }+%
\frac{3(1+\lambda )}{2(1-2\lambda )}H\frac{\dot{\phi}}{\phi }+\frac{%
3(1-\lambda )}{2(1-2\lambda )}p\Bigg \},
\end{equation}%
\begin{eqnarray}
P(H,\frac{dH}{dz},z) &=&\frac{3}{(1+z)}\left\{ \frac{18H_{0}^{2}\biggr \{%
\biggr(\Omega _{m0}(1+z)^{3}+\Omega _{r0}(1+z)^{4}\biggl)\biggr[\frac{%
1-3\lambda }{2(1-2\lambda )}\biggl]+\Omega _{DE}\biggl\}(-1+2\lambda )\phi
^{2}+(-1+3\lambda )\omega \dot{\phi}^{2}}{18H_{0}^{2}\biggr \{\biggr(\Omega
_{m0}(1+z)^{3}+\Omega _{r0}(1+z)^{4}\biggl)\biggr[\frac{1-3\lambda }{%
2(1-2\lambda )}\biggl]+\Omega _{DE}\biggl\}(-1+2\lambda )\phi ^{2}}\right.
\notag \\
&&+\frac{-2\phi ^{2}3H_{0}^{2}\Omega _{m0}(1+z)^{3}+3H_{0}^{2}\Omega
_{r0}(1+z)^{4}+3\lambda \,\phi ^{2}3H_{0}^{2}\Omega
_{m0}(1+z)^{3}+4H_{0}^{2}\Omega _{r0}(1+z)^{4}}{18H_{0}^{2}\biggr \{\biggr(%
\Omega _{m0}(1+z)^{3}+\Omega _{r0}(1+z)^{4}\biggl)\biggr[\frac{1-3\lambda }{%
2(1-2\lambda )}\biggl]+\Omega _{DE}\biggl\}(-1+2\lambda )\phi ^{2}}  \notag
\\
&&\left. -\frac{3H(-2+\lambda )\dot{\phi}+3\lambda \ddot{\phi})}{18H_{0}^{2}%
\biggr \{\biggr(\Omega _{m0}(1+z)^{3}+\Omega _{r0}(1+z)^{4}\biggl)\biggr[%
\frac{1-3\lambda }{2(1-2\lambda )}\biggl]+\Omega _{DE}\biggl\}(-1+2\lambda
)\phi ^{2}}\right\} ,
\end{eqnarray}%
\begin{equation}
Q(H,\frac{dH}{dz},z)=\frac{\omega \dot{\phi}^{2}+\Big(3H_{0}^{2}\Omega
_{m0}(1+z)^{3}+4H_{0}^{2}\Omega _{r0}(1+z)^{4}\Big)-H\phi \,\dot{\phi}+\phi
\,\ddot{\phi}}{2\phi ^{2}\,H_{0}^{2}\biggr \{\biggr(\Omega
_{m0}(1+z)^{3}+\Omega _{r0}(1+z)^{4}\biggl)\biggr[\frac{1-3\lambda }{%
2(1-2\lambda )}\biggl]+\Omega _{DE}\biggl\}(1+z)^{2}},
\end{equation}%
where $\dot{\phi}=\phi _{0}\sigma \,t^{\sigma -1}$ and $\ddot{\phi}=\phi
_{0}\,\sigma \,(\sigma -1)\,t^{\sigma -2}$.

\begin{figure}[h]
\centering
\begin{tabular}{rl}
\includegraphics[width=9cm, height=7cm]{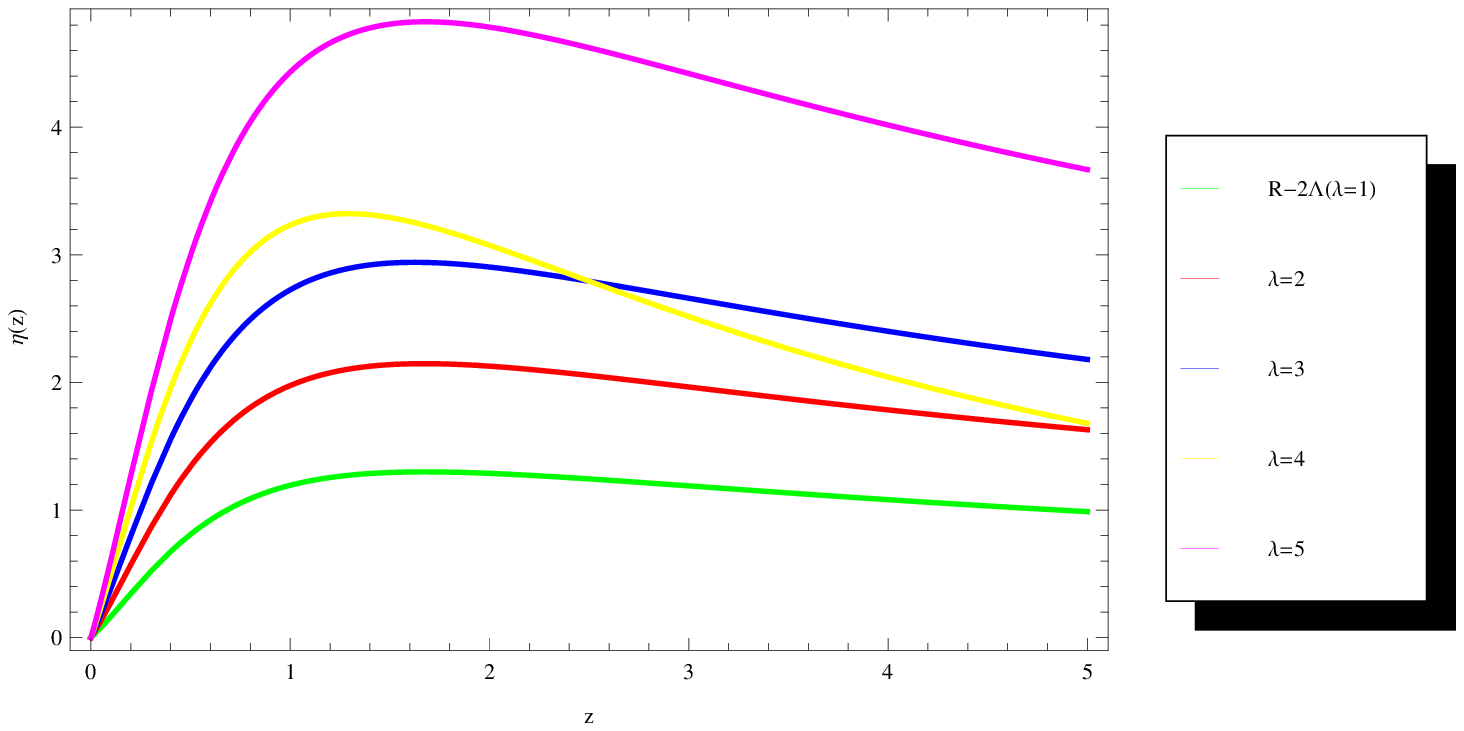} & %
\includegraphics[width=9cm, height=7cm]{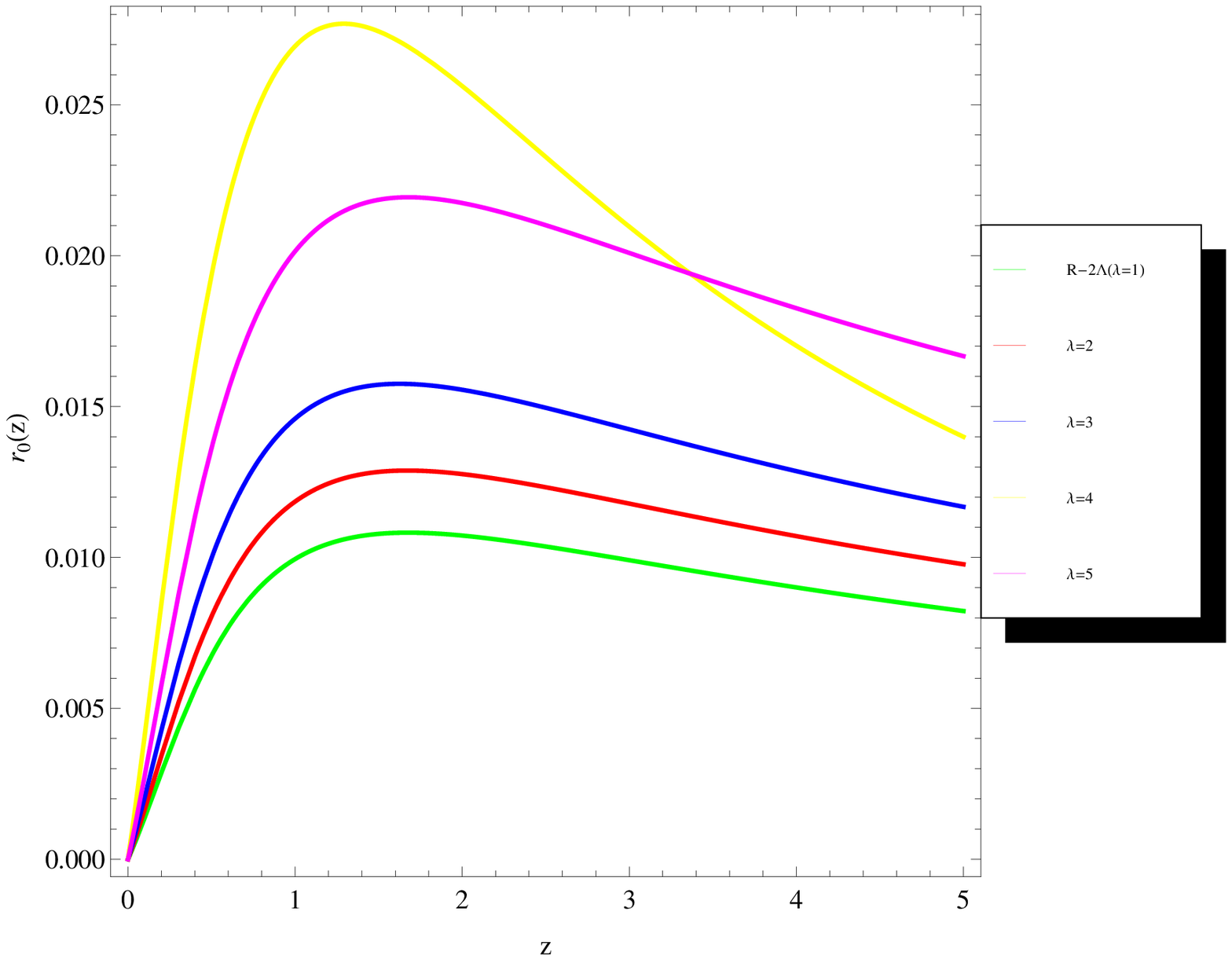}%
\end{tabular}%
\caption{The graphs shows the deviation vector magnitude $\protect\eta (z)$
(left panel) and observer area distance $r_{0}(z)$ (right panel) for null
vector field GDE with FLRW background as functions of redshift. The graphs
are plotted for $H_{0}=80Km/s/Mpc$, $\Omega _{m0}=0.3$, $\Omega _{r0}=\Omega
_{k0}=0$, $\Lambda =1.7.10^{-121}$ and we imposed in equation (\protect\ref%
{salako57}) the initial conditions $\protect\eta (z=0)=0$ and $\protect\eta %
^{\prime }(z=0)=1$.}
\label{fig3}
\end{figure}

We numerically solve the differential equations (\ref{salako67})
($\Lambda CDM$) and (\ref{MattigGen'}) (BDR) by plotting  the
deviation vector magnitude $\protect\eta $ and observer area
distance $r_{0}$ as functions of redshift $z$ (Fig. \ref{fig3}). For
physical reasons, we will choose the values of $ \lambda $ as in
\cite{Carames, Carames'}. In each panel of (Fig. \ref{fig3}), we
observe that the evolution of the deviation $\eta(z)$ and observer
area distance $r_0(z)$ show similar behavior to those of
$\Lambda$CDM. Within this model, we can see as the
$\lambda\geqslant1 $ increases and when one goes to the highest
values of the redshift $(z \leqslant 0.5)$, the deviation $\eta(z)$
and observer area distance $r_0(z)$ decouple each of the model
$\Lambda$CDM but still keeps the same place, while for the low
values of the redshifts, i.e., for the current day, the BDR model
reproduce exactly $\Lambda$CDM. We can conclude that the results are
similar to $\Lambda$CDM for all the cases, which means that the
above BDR model can be considered remain phenomenologically viable
and tested with observational data. We can also observe for small
value of redshift $0< z <0.5$, the magnitude of deviation vector
$\protect\eta $ presents the same behavior that of $\Lambda CDM$
model ($\lambda=1$). Also, the same behavior was observed for area
distance $r_{0}$ at the same level. However, when considering large
values of redshift $z > 0.5$, we find a gap between the BDR model
($\lambda\neq1$) and the $\Lambda CDM$ model. This indicates that
the BDR model predicts a strong acceleration than $\Lambda CDM$
model for large values of redshift.

\section{Conclusions}

In this paper, we have considered GDE in BDR gravity and calculated
the Ricci tensor and Ricci scalar with the modified field equations
in BDR gravity. Then, in FLRW universe, corresponding GDE for BDR
gravity is obtained. We restricted our attention to extract the GDE
for two special cases, namely the fundamental observers and past
directed null vector fields. In these two cases, we have found the
Raychaudhuri equation,  GDE for null vector fields and the
diametral angular distance differential for BDR gravity.  We
have also numerically computed the geodesic deviation ($\eta(z)$)
and the area distance ($r_{0}(z)$) from Mattig's relation for BDR
gravity models and compared our results with those of $\Lambda $CDM
model (Fig.\ref{fig3}).

\acknowledgments{Ines G. Salako and M. J. S. Houndjo  thank IMSP for
hospitality during the elaboration of this work.}

\end{document}